# A Methodology for Deriving Evaluation Criteria for Software Solutions

Harald Papp[1][1], Marc Hanussek[1][2]

[1]Department for Digital Business, University of Stuttgart, Institute of Human Factors and Technology Management (IAT)
Nobelstraße 12, 70569 Stuttgart, Germany
{harald.papp, marc.hanussek}@iat.uni-stuttgart.de.

Abstract: Finding a suited software solution for a company poses a resource-intensive task in an ever-widening market. Software should solve the technical task at hand as perfectly as possible and, at the same time, match the company strategy. Based on these two dimensions, domain knowledge and industry context, we propose a methodology for deriving individually tailored evaluation criteria for software solutions to make them assessable. The approach is formalized as a three-layer model, that ensures the encoding of said dimensions, where each layer holds a more refined and individualized criteria list, starting from a general software-agnostic catalogue we composed. Finally, we exemplarily demonstrate our method for Machine-Learning-as-a-Service platforms (MaaS) for small and medium-sized enterprises (SME).

## 1 INTRODUCTION

Increasing digitization offers huge potential for enterprises to streamline and automate processes and services and thus increase the value creation in the company (Loebbecke & Picot, 2015). Choosing the right software solution for internal processes is a crucial step towards optimal workflow. This requires not only a constantly updated market overview to catch up on the ever-increasing software-supply, but an accurate assessment of the company needs (Schmidt, et al., 2015).

However, the variety of services and the specificity of solutions makes it difficult for non-experts to choose the right product. On the other hand, technical experts might not have the strategic insight, as well as detailed industry knowledge.

For orderly assessment of such challenges, requirements managers are employed. Their scope of action includes leading the communication between internal and external stakeholders, transforming overall objectives into tangible requirements. They aim to generate a mutual understanding of the complex problem at hand between all involved parties (Stellman & Greene, 2005).

Since the selection of a software solution for company-internal purposes neglects external stakeholders, we will address a subset of common requirements management methodologies. In this paper we hypothesize that the problem space for choosing an optimal company-intern software solution is comprised of two dimensions.

Accordingly, a requirements manager in the context of this work is not to be understood from a common project management perspective, but in the role of flexibly adopting the two mentioned dimensions: domain knowledge and industry context.

### 1.1 Domain Knowledge vs. Industry Context

For the scope of this paper we define two abstract dimensions to describe the theoretical basis of our approach. This simplifies a multi-dimensional problem to a more manageable setting. Every possible aspect is then part of either one of the dimensions.

---

[1] https://orcid.org/0000-0002-8255-1556
[2] https://orcid.org/0000-0002-8041-8858

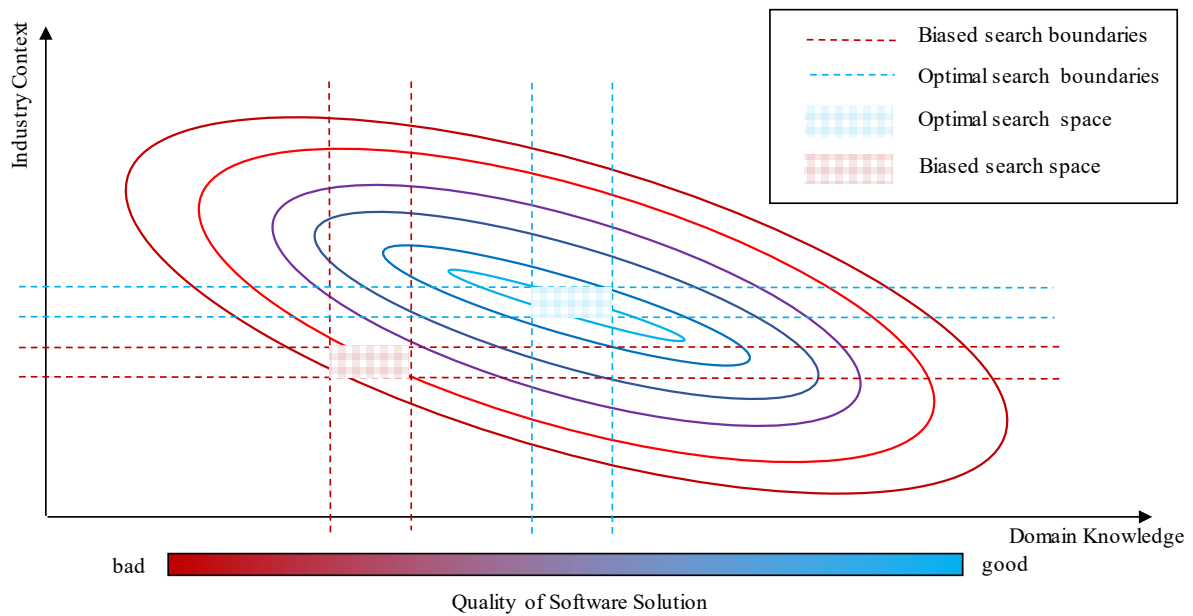

Figure 1: Schematic depiction of the theoretical foundation of the model. On the axes, the two presented dimensions are depicted (although not necessarily orthogonal). A point on the graph is defined as the matching score: How well is a software solution suited. Theoretical matching scores for a specific software solution are illustrated in form of an ellipse.

There is domain knowledge, the knowledge coming from an expert of the field, where the new software solution is to be employed. The expert works with the current suboptimal product, can assess its perks and drawbacks, knows about the future professional trajectory of his field and thus is able to generate requirements from his perspective.

On the other hand, the term industry context summarizes all information about company size, market segmentation and niche of the company as well as its strategy. It is a broader, coarser perspective, inspired by KPIs (Key Performance Indicators) and company policies, like interfaces, personal resource allocations together with company guidelines.

## 1.2 Different Actors Propose Different Criteria

These two dimensions must not be understood as mathematical dimensions, where linear independence holds, but from a pragmatic company-hierarchy perspective. They depict the perspective of different states of mind from individual actors in a company and both need to be encoded into a criteria list for a software solution, so that all requirements can be met.

If an actor, who thinks in industry context terms proposes a list of criteria for assessing the optimal software solution, simply by definition of human nature, he encodes a bias to the criteria list (Evans, 1989). For example, low cost and low implementation effort might be driving factors.

This holds for an actor, who thinks in domain knowledge terms, vice versa. Here, the bias might be directed towards performance and functionality.

## 1.3 Choosing the Wrong Search Space

Thus, a criteria list which stems from a single party tends to be incomplete. The so-created search space in the field of all possible software solutions, does not necessarily allow for a global maximum (ergo the best possible solution), but tends to result in a local one (a solution which is subjectively optimal). Only by bringing both dimensions into account, the optimal solution can be found in the overlap of the respective search spaces (see Figure 1). De facto, especially large companies tend to struggle to bring the two dimensions together (Lund & Gjerding, 1996).

However, there is a wide spectrum of proposed methodologies to derive criteria for assessing the best software solution, that try to solve such problems.

## 1.4 Related Work

In (Jadhav & Sonar, 2009) the authors review evaluation and selection of software packages. They discuss various software evaluation techniques such as Analytic Hierarchy Process, feature analysis, weighted average sum and a fuzzy based approach. They also provide evaluation criteria which they subdivide into categories such as Functional, Personalizability, Portability, Maintainability, Usability, Reliability, Efficiency, Vendor, Cost and Benefits. The authors clearly state that "there is lack of a common list of generic software evaluation criteria and its meaning".

In (Godse & Mulik, 2009) selection of Software-as-a-Service Product is discussed. They use the Analytic Hierarchy Process for calculating weights of selection parameters and scores for products. The selection parameters they quote, are Functionality, Architecture, Usability, Vendor Reputation, and Cost. They assert that "there is no explicit guidance available on selection of SaaS product for business application".

The authors of (Ekanayaka, Currie, & Seltsikas, 2003) evaluate application service providers (ASP) by using an evaluation framework that comprises categories such as "security, pricing, integration, service level agreement, and reliability, availability and scalability". They state that SMEs with limited experience of IT outsourcing can "enter the ASP market at reduced risk as long as they learn to evaluate disparate ASP offerings". At that time, it was too early to assess the success of ASPs, which is also stated in the paper. Nowadays we (continue to) observe rapid growth in the Software-as-a-service (SaaS) market. While the ASP model and SaaS are not the same thing, the basic concept is resembling and the need for well-defined specific evaluation frameworks or criteria remains high.

Hence, we propose a three-layer method for deriving a requirements criteria catalogue for software products, tailored not only for specific use cases, but also accounting for the industry context of the company. Applying this catalogue to different software solutions gives a rating score per solution and allows comparison.

## 2 METHODS

Our method is comprised of three layers, each a list of criteria and two connections, defining the transitions between the layers (Figure 2).

In a first step a list of generic requirements criteria for software products is refined to a domain-specific subset by employing domain knowledge. The second step then weighs the criteria in the subset to mirror industry context. The resulting list of criteria then allows to rate a software solution on every criterion with a numeric value. Adding all values then results in a Matching Score (MS) that reflects both:
- How well does the software solution solve the technical problem a business has?
- How well does the software solution line up with the business strategy?

While we will exemplarily apply this approach to Machine Learning as a Service (MaaS) for small and medium-sized enterprises in Section 3, the method should allow for an employment in overall software solutions.

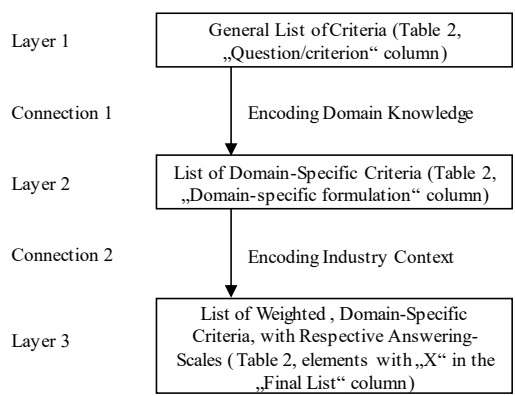

Figure 2: Illustration of the three-layer methodology to derive a criteria catalogue for software solutions

### 2.1 Layer 1 – List of Criteria for General Software Solutions

The first layer of our method is a list of use case and industry agnostic criteria formulated as questions. Some are formulated in such a way, that they can be answered on a Likert scale[3], others can be answered by numeric values. Scaling will become relevant in the second layer-connection.

Every criteria list-element belongs to a different category. Table 1 shows a brief overview of those categories, that hold most questions.

The list is partly composed from ISO Norms (ISO, 2011), as well as different requirements models (Jadhav & Sonar, 2009), (Ekanayaka, Currie, & Seltsikas, 2003), (Brand, 2017), (Godse & Mulik, 2009), (SoftGuide GmbH & Co. KG, 2020), (Ludewig, 2011). Semantic overlap between different models was resolved, by summarizing similar questions from different models into one element.

---

[3] A Likert scale (Likert, 1932) is the most widely used approach for measuring personal opinions. The typical five-level form of the Likert scale consists of the five manifestations *Strongly disagree, Disagree, Neither agree nor disagree, Agree* and *Strongly agree.*

In total, the list has 62 elements and tries to give an as complete and generic as possible criteria catalogue for software solutions. The complete list can be found in the Appendix (Table 2, "Question/criterion" column).

Table 1: Overview of criteria categories

| Category | Number of Elements in Category |
|---|---|
| Usability | 19 |
| Documentation and support for different languages | 7 |
| Costs | 3 |
| Performance | 3 |
| Requirement of workers and their skill | 4 |

## 2.2 Layer 2 – Deriving a Domain-Specific Criteria List

The second layer results in a list of domain specific criteria or in other words, the search space should be correctly localized in the domain knowledge dimension (see Fig. 1). To assess the correct search space boundaries, the connection from the first layer (list of criteria for general software solutions) to the second layer is to be specified:

Encoding domain specific knowledge and information from experts of the field to the list of general criteria is key. This is done in two steps:
- Identify those criteria, that are impractical or cannot be applied to the software solution under scrutiny. Remove them.
- Reformulate the remaining criteria to domain-specific wording. Scaling is irrelevant at this stage.

## 2.3 Layer 3 – Deriving a Weighted Domain-Specific Criteria List

The final, third layer needs to have industry context encoded into it. Further, it depicts the final list of criteria that makes a software solution rate-able and therefore comparable.

For this matter, a wide spectrum of methods, spanning from analytic hierarchy processes to fuzzy based approaches, is available. In (Jadhav & Sonar, 2009) a weight-based approach (weight average sum (WAS)) was called to be the easiest to use. As a downside, the arbitrariness of assigning weight-values to criteria list-elements was criticized. We therefore propose a corrected weight-based approach, accompanied by a rule set to induce more objectivity into the weighting process. The connection from layer 2 to 3 is defined as follows:
- Before assigning weight-values, every criteria list-element in layer 2 is rated based on importance to the business, business-strategy, etc.: an all-over industry context assessment, where a high number shows significance to the business and a low number insignificance. We found that a 1 to 5 scale was sufficiently fine-grained, but coarse enough to be distinct. However, differently numbered scales are also possible.
- Every element from layer 2 is examined, whether it represents a showstopper to the business.
- Depending on the rating and on being a showstopper or not, every element is assigned a scale (Boolean, Likert, Numerical). Based on this scale, a specific criterion is later rated and reflected in the final assessment of a software solution. The rule set to match a scale to an element is defined as follows (Further above rules overrule lower rules):
  - Element is a showstopper: Boolean
  - Element is rated 1-3: Boolean
  - Element is rated 4,5: Likert
  - Element specifically asks for a numeric value: Numeric

Thus, missing showstopper criteria impose a high penalty, because they simply won't appear in the MS. Important (4, 5), but non-showstopper criteria leave space to accurately weigh them, while unimportant criteria (1, 2, 3) are not taken into account in such detail and being Boolean, can be "turned" on and off, reducing the noise in the MS.

- Every criteria list-element is reformulated in such a way, that it can be answered on the assessed scale.
- Weights are normalised and depicted as percentages.

## 2.4 Giving Matching Scores to Software Solutions

Layer 3 is comprised of ***K*** Numeric-scaled criteria ($d_k \in \mathbb{R}^+$), ***L*** Boolean-scaled criteria ($b_l \in \{0,1\}$)

and $M$ Likert-scaled criteria ($c_m \in \{1,2,3,4,5\}$) and thus total in $N$ criteria, so that

$$N = K + L + M \quad (1)$$

holds. A software solution is then respectively rated in every criteria list-element with normalised[4] Numeric values ($a_k^{Numeric}$), Boolean values ($a_l^{Boolean}$) and Likert-values ($a_m^{Likert}$). The Matching Score (MS) is:

$$\text{MS} = \sum_0^L a_l^{Boolean} b_l + \sum_0^M a_m^{Likert} c_m + \sum_0^K a_k^{Numeric} d_k \quad (2)$$

## 3 APPLICATION: MAAS PLATFORMS CRITERIA FOR SME

This section exemplarily applies the methodology described in Section 2 to a scenario in which a SME is in search for a MaaS solution, hence intending to prepare a selection process. This scenario is intentionally abstract in order to address as many concerned parties as possible. Followingly, we establish a minimum of presumptions.

Beginning with the complete list (see Table 2, "Question/criterion" column) criteria which are not relevant for the specific domain are removed. Here, the domain is Machine learning as a Service, which can be understood as the overlap of machine learning on the one hand and cloud services on the other hand. While there are several reasons that lead to removal of criteria, we describe the three main ideas in the following. The complete list of removed criteria, complemented by the respective reason, can be found in Table 2 (elements with no "X" in the "Final List?" column).

The first reason is that, due to its nature, cloud solutions do not need to be installed on a local computer or connected to the corporate network. More specifically, MaaS solutions usually run in server clusters which are hosted by the vendor. The customer can comfortably access the service via web browser or APIs. Consequently, criteria such as 4.1 and 22.4 are removed.

Another reason for removal concerns servicing and maintenance aspects. Normally, the vendor of MaaS platforms manages all arising technical issues such as (security) incidents, customer support, preventive maintenance and updating software. Ideally, the user is not even aware of these operations and can fully concentrate on his business use case and its implementation in the platform. This leads to removal of, for example, criteria 9.1 and 20.2.

The third reason deals with the fashion MaaS solutions are used. Usually, the products are offered in a highly service-oriented fashion. Next to servicing and maintenance aspects, this particularly applies to modification of the solution. Normally, vendors of MaaS solutions provide complete services which do not need to be customized by the customer. Vendors strive to offer generic interfaces such that it is not necessary (or possible, or desired, respectively) for users or other agents to adjust the software to users' needs. Consequently, structure, maintainability and documentation of source code is not relevant for the customer, at least not immediately. These are, among others, criteria 22.6 and 22.7.

After the removal process, we face a list of criteria which is relevant for MaaS solutions (see Table 2, elements with "X" in the "Final List?" column). For the next step, that is refinement to specific target audiences, we look into specific needs and customs of the considered company or its industry. Concretely, consider a SME whose primary field of activity is not (machine learning centered) IT. By weighting the remaining criteria, we aim to obtain a final list of criteria of varying importance. Similar to the previous step, we subsequently present the main reasons and ideas behind this exemplary approach. Note that, while different companies and business sectors can have immensely differing needs and customs, we try to focus on possible similarities. We hereby invoke experiences from our project work in diverse fields of application.

The first assumption is lack of machine learning specialists in the company. This hypothesis should hold true for most of the concerned businesses, since we consider companies that do not primarily operate in (machine learning centered) IT by assumption. Furthermore, hiring such staff is hard, as the hype for machine learning is a relatively new phenomenon, hence limiting the number of graduates. Having the aforementioned lack in mind, we assume that the considered company will conduct comparatively basic experiments on the MaaS platform. Consequently, expert features of the MaaS solution like 2.19, 6.2 and 13.1 are not of particular importance and will be assigned minor weights.

---

[4] High numerical values bias the Matching Score: although a numeric criterion might be weighted with a 1, following formula (2) it strongly outweighs even showstopper-criteria. This is not negatively reflected in the comparison between software solutions, because every numeric value is on the same scale across the different solutions, however a normalization of numeric values to a proper range, ensures the consistency within a single software solution.

Limited resources are another impact factor for the evaluation. We suppose that SMEs neither have sufficient reserve assets, nor enough staff for distinct advance development. Instead, SMEs need decent return on investment in a comparatively short time period. Hence, the MaaS solution should either be low in price or generate quick net product. Criteria fulfilling these requirements will be assigned major weights (e.g. 3.2, 3.3).

Lastly, like every other company, SMEs must deal with showstoppers. Such criteria will be assigned the highest possible weight. Apart from universal showstoppers, like non-compliance with locally applicable law, we identified one noteworthy MaaS-specific criterion: reliability or the correctness of the system's output (15.1). In machine learning, estimating model performance for unseen data is a complex task. Small or biased datasets can complicate this task even more. Combined with non-expert users, this bears the pitfall of overestimating model performance which, in turn, can have economic consequences. Next to mere model performance in terms of known evaluation metrics, there are additional risks such as estimators learning side issues instead of focusing on relevant aspects of the data. In image classification, for example, there are known cases in which this leads to unexpected behavior of image classifiers (Han S. Lee, 2017). Therefore, in our scenario it is utterly important for MaaS solutions to feature robust model performance estimates such as cross validation on the one hand and provide model insight (possibly with methods of the field of explainable artificial intelligence) on the other hand. Thereby, the risk of misuse by non-expert users can be reduced.

The three aforementioned assumptions, together with additional considerations (see again Table 2), lead to the weights column. By normalizing, as described in Section 2.3, we obtain the final list exhibiting the most import criteria which is adjusted to the domain MaaS as well as to the target audience SMEs.

## 4 CONCLUSIONS

We presented a transferable methodology for deriving a criteria catalogue for software solutions. It can be directly applied as is or used as an inspiration for problems alike. As such, different software solutions (of the same scope) on the market can objectively be compared, so that the optimal solution can be found for a business.

To allow for this, we proposed that two independent dimensions – domain knowledge and industry context – need to be encoded into a template criteria catalogue (which we also compiled). The first dimension ensures, that the software solution indeed solves the technical problem one faces, the second dimension attests, that it is in line with business strategy and branch context. Followingly, one could consider domain knowledge, as a bottom-up process, reflecting the skill of specialists, while industry context mirrors a top-down process, reflecting the market understanding of decision makers.

The method is formalized in a three-layer model with two layer connections in between. Because the first layer is a general list of criteria, blended together from many sources, it should offer a software-agnostic basis for tackling decision problems. The transition to the second layer encodes domain knowledge and the connection to the third layer encodes industry context. Connections were presented as a step sequence, accompanied by examples, to additionally illustrate the approach.

The resulting catalogue in layer three can then be used for the comparison of software solutions: Every software solution under scrutiny, is assessed in every criterion, yielding a final matching score.

We plan to further evaluate the theoretically derived model in an empirical study.

## REFERENCES

Brand, K. (2017, April 27). *www.xing.com*. Retrieved from https://www.xing.com/news/insiders/articles/wesentliche-kriterien-fur-die-auswahl-von-it-losungen-705949

Ekanayaka, Y., Currie, W., & Seltsikas, P. (2003). Evaluating application service providers. *Benchmarking*.

Evans, J. (1989). *Bias in human reasoning: Causes and consequences.* Lawrence Erlbaum Associates, Inc.

Godse, M., & Mulik, S. (2009). An Approach for Selecting Software-as-a-Service (SaaS)

Product. *2009 IEE International Conference on Cloud Computing*.

Han S. Lee, A. A. (2017, September). Why Do Deep Neural Networks Still Not Recognize These Images?: A Qualitative Analysis on Failure Cases of ImageNet Classification. Retrieved from https://arxiv.org/abs/1709.03439

ISO, I. (2011). Iec25010: 2011 systems and software engineering--systems and software quality requirements and evaluation (square)--system and software quality models. *International Organization for Standardization*, p. 2910.

Jadhav, A. S., & Sonar, R. M. (2009). Evaluating and selecting software packages: A review. *Information and Software Technology*.

Likert, R. (1932). A technique for the measurement of attitudes. *Archives of Psychology*, pp. 22 140, 55. Retrieved from https://legacy.voteview.com/pdf/Likert_1932.pdf

Loebbecke, C., & Picot, A. (2015). Reflections on societal and business model transformation arising from digitization and big data analytics: A research agenda. *The Journal of Strategic Information Systems 24.3*, pp. 149-157.

Ludewig, E. (2011). *https://www.usabilityblog.de*. Retrieved from https://www.usabilityblog.de/messen-zahlen-vergleich-%E2%80%93-diese-metriken-konnen-usability-tests-bereichern/

Lund, R., & Gjerding, A. (1996). The flexible company: innovation, work organization and human resource management. *Danish Research Unit for Industrial Dynamics (DRUID) Working Paper*.

Schmidt, R., Zimmermann, A., Möhring, M., Nurcan, S., Keller, B., & Bär, F. (2015). Digitization–perspectives for conceptualization. *European Conference on Service-Oriented and Cloud Computing*.

SoftGuide GmbH & Co. KG. (2020). *https://www.softguide.de*. Retrieved from https://www.softguide.de/software-tipps/

Stellman, A., & Greene, J. (2005). *Applied software project management.* O'Reilly Media, Inc.

# APPENDIX

Table 2: Complete list of criteria, as well as the refined criteria for the example in section 3. The "Question/criterion" column depicts the full list of general criteria for software solutions. If an element is relevant for the example in section 3 it is marked with an "X" in the "Final List?" column. It is reformulated in the "Domain-specific formulation" column and rated, weighted and fitted with a scale ("Rating, Weighting, Scale" column). The "Justification" column bears a description for the elimination from the list for the example or for the rating for the example.

| Index Category | Question/criterion | Domain-specific formulation | Final List? | Justification | Rating Weighting Scale |
|---|---|---|---|---|---|
| 1.1 Functionality | What added value does the IT solution bring to the business? | What added value does the IT solution bring to the business? | X | SMEs do not have the resources to experiment on a long-term basis (showstopper). | 5 (3,8%) boolean |
| 1.2 Functionality | What is the time to availability? | How long does it take for compute resources, trained models, or predict requests to be available or processed? | X | SMEs do not primarily use MaaS for time-critical applications. | 2 (1,5%) intervals |
| 2.1 Usability | Is the user interface intuitive? | Is the user interface intuitive? | X | Non-experts need intuitive UI. | 4 (3,0%) Likert |
| 2.2 Usability | Does the dialog only show user information related to the completion of the work item? | Does the dialog only show user information related to the completion of the work item? | X | Non-experts are otherwise overwhelmed. | 3 (2,3%) boolean |
| 2.3 Usability | How helpful is contextual help? | How helpful is contextual help? | X | Non-experts need help. | 4 (3,0%) Likert |
| 2.4 Usability | What is the training effort? | How-time consuming is it to apply the first ML models? | X | SMEs do not have the resources to experiment on a long-term basis. | 4 (3,0%) Likert |
| 2.5 Usability | Is there support for recurring tasks (e.g. macros; in MaaS context: pipeline)? | Can recurring ML flows be stored in pipelines and run repeatedly? | X | Rare use of complex pipelines. | 1 (0,8%) boolean |
| 2.6 Usability | Are there undo-features? | Are there undo-features? | X | It is customary and user-friendly to have this option. | 1 (0,8%) boolean |
| 2.7 Usability | Self-description capability: How useful and instructive is feedback? | Self-description capability: How useful and instructive is feedback? | X | Non-experts need help. | 4 (3,0%) Likert |
| 2.8 Usability | Self-description capability: Are there further inquiries for important operations? | Self-description capability: Are there further inquiries for important operations? | X | Non-experts are not as experienced as experts when it comes to serious decisions. | 3 (2,3%) boolean |
| 2.9 Usability | Is it possible to resume at the starting point after interruption? | | | In cloud context, it is typical for IT systems or services to run nearly perpetually. | |
| 2.10 Usability | Is it possible to recover last deleted objects? | Can deleted workflows or ML models be restored? | X | Rare use of complex pipelines. | 1 (0,8%) boolean |
| 2.11 Usability | Expectation conformity: Are the comprehension requirements of the dialog consistent with the user's knowledge? | Expectation conformity: Are the comprehension requirements of the dialog consistent with the user's knowledge? | X | Overchallenged users are likely to make mistakes. | 3 (2,3%) boolean |
| 2.12 Usability | Expectation conformity: Is vocabulary used the user is familiar with? | Expectation conformity: Is vocabulary used the user is familiar with? | X | Familiar vocabulary is user-friendly and reduces the risk of maloperation. | 3 (2,3%) boolean |
| 2.13 Usability | Conformity of expectations: Are dialogues for similar work tasks designed similarly? | | | In our opinion, helpful tooltips (2.3) provide better assistance. | |
| 2.14 Usability | Expectation conformity: Do system responses occur immediately? | Expectation conformity: Do system responses occur immediately? | X | This allows for immediate correction of incorrect user input. | 3 (2,3%) boolean |
| 2.15 Usability | Fault tolerance: How useful are display and explanations of input errors? | Fault tolerance: How useful are display and explanations of input errors? | X | Non-experts need help. | 4 (3,0%) Likert |
| 2.16 Usability | Fault tolerance: Is input data checked for validity and confirmed before use? | To what extent does the service check the incoming data for | X | Non-experts need help. | 4 (3,0%) Likert |
| | | compatibility with the training phase? | | | |
| 2.17 Usability | Customizability: Are settings adapted to specific needs and capabilities of the user? | Is prior knowledge of machine learning adequately addressed? | X | Typically, considered SMEs do not have multiple users with varying knowledge of machine learning. | 1 (0,8%) boolean |
| 2.18 Usability | Customizability: Is adaptation to language, knowledge, cultural peculiarities (e.g. key-binding), motoric skills and perceptual capacity of the user possible? | | | Machine learning is a field which is deeply penetrated by English language and conventions. | |
| 2.19 Usability | Customizability: Can output be presented individually? | Can results be displayed differently, for example by different error measures? | X | Too detailed for the beginner. | 2 (1,5%) boolean |
| 3.1 Costs | What are one-time vs. ongoing costs? | What are one-time vs. ongoing costs? | X | Generally, SMEs are cost-conscious. This not only applies to one-time costs but also to ongoing costs. | 4 (3,0%) numeric |
| 3.2 Costs | What is the total cost of ownership (TCO) for the IT solution? | What is the total cost of ownership (TCO) for the IT solution? | X | Limited budget. | 4 (3,0%) numeric |
| 3.3 Costs | What is the near-term vs. long-term Return of Investment? | What is the near-term vs. long-term Return of Investment? | X | Financial lean period due to limited reserves inappropriate. | 4 (3,0%) intervals |
| 4.1 Performance of the IT solution | Does the IT solution run at decent speed on standard local hardware? | | | No installation of software needed. | |
| 4.2 Performance of the IT solution | What is the handling time? | How much time does the training of the models take? | X | SMEs do not primarily use MaaS for time-critical applications. | 2 (1,5%) intervals |
| 4.3 Performance of the IT solution | Reliability: Does the system deliver correct results? | How is ensured that the system learns the right thing? | X | Non-experts otherwise overestimate the capabilities of the application. | 5 (3,8%) Likert |
| 5.1 Requirement of manpower and knowledge/ability | Does the IT solution require a high level of manpower (including rare knowledge/skills)? | | | In MaaS context, users can decide on their own, how much manpower they want to put into the tool. | |
| 6.1 Scalability | Can the IT solution increase its output by adding additional resources (typically hardware) to handle the increased load? | Can the service scale hardware sufficiently to create more powerful models or handle more predict calls? | X | Data volume does not change dramatically in SMEs over time. | 2 (1,5%) boolean |
| 6.2 Scalability | Further development: Can the existing solution be further developed? | Can trained models be refined manually? | X | No ML experts available in SMEs. | 1 (0,8%) boolean |
| 6.3 Scalability | Testability: What is the effort required to test the modified software? | | | Updates and patches in cloud environment are backwards compatible and automatically installed on service side. | |
| 7.1 Agility, flexibility, adaptability | Can the IT solution be easily and quickly adapted to new requirements (e.g. without programming)? | Can the IT solution be easily and quickly adapted to new requirements, such as more data points or attributes? | X | Data type does not change dramatically in SMEs over time. | 2 (1,5%) boolean |
| 7.2 Agility, flexibility, adaptability | Modifiability: What is the effort to perform improvements, troubleshooting, or adapt to environmental changes? | | | Elimination of errors and improvements in cloud environment are automatically carried out on service side. | |
| 8.1 Modularity | Does the IT solution have a modular or monolithic architecture? | | | Users do not maintain or extend the product. | |
| 9.1 Serviceability | Is it easy to install, operate, maintain, and upgrade the IT solution? | | | Maintenance, upgrades and running are performed as a service by the cloud service. | |
| 10.1 Portability | Is it possible to transfer the software to another system environment? | | | MaaS tools are accessed through a web browser, hence no transfer of software is needed. | |
| 11.1 Interfaces | Does the IT solution offer open or proprietary interfaces to connect to other IT solutions? | To what extent does the IT solution provide open or proprietary interfaces to read data or receive predict calls? | X | Interfaces make the MaaS application user-friendly. | 4 (3,0%) Likert |
| 11.2 Interfaces | Can machine learning models be exported? | Can machine learning models be exported? | X | No ML experts available in SMEs that maintain models locally. | 1 (0,8%) boolean |
| 12.1 Interoperability | Can the IT solution work easily smoothly with other IT solutions (e.g. through standard interfaces and data models)? | Can the IT solution work easily smoothly with other IT solutions (e.g. through standard interfaces and data models)? | X | Own interface development is too complex. | 4 (3,0%) Likert |
| 13.1 Multi-client capability | Does the IT solution offer the ability to set up multiple clients (such as company codes) that can run independently? | Is it possible to create multiple parallel ML workflows and/or train models independently of each other at the same time? | X | No ML experts available in SMEs. | 1 (0,8%) boolean |
| 14.1 Cloud capability | Can the IT solution be operated as a private or public cloud | | | Obviously, we consider only cloud services. | |
| | solution (Software as a Service, Platform as a Service)? | | | | |
| 15.1 Maturity, reliability, fault tolerance | How mature, reliable, or fault-tolerant is the IT solution (e.g. restart without data loss after failure)? | How mature, reliable, or fault-tolerant is the IT solution (e.g. restart without data loss after failure)? | X | No resources to deal with ever-changing platform conditions. | 5 (3,8%) Likert |
| 15.2 Maturity, reliability, fault tolerance | How proven is the software in the short-term vs in the long term? | | | MaaS solutions are in continuous change and brisk adoption by practitioners was not long ago. | |
| 16.1 Sustainability | Will the IT solution be further developed and supported by the IT solution provider in the medium to long term? | Are new ML features like new model types added? | X | Standard procedures are sufficient. | 1 (0,8%) boolean |
| 16.2 Sustainability | How frequency are there updates? | To what extent is support provided? | X | Support cannot be provided independently. | 4 (3,0%) Likert |
| 16.3 Sustainability | Are new features or bug fixes implemented? | Are new features or bug fixes implemented? | X | Standard procedures are sufficient. | 1 (0,8%) boolean |
| 17.1 Compliance with enterprise IT architecture | Does the IT solution meet the standards set by your organization's enterprise IT architecture? | | | The service is performed in an environment outside the company. | |
| 18.1 Compliance with laws and regulations | Does the IT solution meet all relevant legal and regulatory requirements (e.g. principles of sound accounting)? | Does the IT solution meet all relevant legal and regulatory requirements (e.g. principles of sound accounting)? | X | Applies to any company (showstopper). | 5 (3,8%) boolean |
| 19.1 IT governance | Does the IT solution adequately support IT governance requirements? | Does the IT solution adequately support IT governance? | X | Even SMEs need basic IT governance aspects. | 2 (1,5%) boolean |
| 20.1 Information security | Does the IT solution's information security architecture provide adequate protection against information security threats? | Does the IT solution's information security architecture provide adequate protection against information security threats? | X | Applies to any company (showstopper). | 5 (3,8%) boolean |
| 20.2 Information security | Analysability: What is the effort required to diagnose defects or causes of failure or to determine parts in need of change? | | | Troubleshooting and maintenance are performed as a service by the cloud service. | |
| 20.3 Information security | Are there versioning features and historical views of the data? | Are there versioning features and historical views of the data? | X | Standard settings are sufficient. | 1 (0,8%) boolean |

| | | | | | | |
|---|---|---|---|---|---|---|
| 21.1 Data privacy | Does the IT solution adequately protect corporate data (personal data, customer data, intellectual property)? | Does the IT solution adequately protect corporate data (personal data, customer data, intellectual property)? | X | Applies to any company (showstopper). | 5 (3,8%) boolean |
| 22.1 Documentation and support for different languages | How good is the documentation of the IT solution for users and operators? In which languages is the documentation available? | How good is the documentation of the IT solution for users and operators? In which languages is the documentation available? | X | Non-expert needs good documentation. | 4 (3,0%) Likert |
| 22.2 Documentation and support for different languages | Is there a programmer documentation (description of source code)? | | | The majority of MaaS providers are private businesses operating with a view to gain, thus keeping source code private. | |
| 22.3 Documentation and support for different languages | Method documentation: Are mathematical algorithms, technical-scientific or commercial methods properly described? | How well are the methods used (e.g. cross-validation) and AI algorithms or their results described and explained? | X | Non-expert needs help. | 4 (3,0%) Likert |
| 22.4 Documentation and support for different languages | Is required hardware, software, possible operating systems, standard libraries or runtime systems, installation, updates and deinstallation properly described? | | | No installation needed since this is performed as a service by the cloud service. | |
| 22.5 Documentation and support for different languages | Is there a data documentation (formats, data types, restrictions, import and export interfaces)? | Is there a data documentation (formats, data types, restrictions, import and export interfaces)? | X | Non-expert needs help. | 4 (3,0%) Likert |
| 22.6 Documentation and support for different languages | Is there a test documentation? | | | Users generally are not concerned with software tests. | |
| 22.7 Documentation and support for different languages | Is there a development documentation? | | | Users generally are not integrated in the development process. | |
| 23.1 Innovative character | How common is the solution in the market? | How common is the solution in the market? | X | Users could get help in online communities if solution is widely used in market. | 3 (2,3%) Likert |
| 24.1 Manufacturer dependency | Does using the solution make you tied to a single manufacturer? | Does using the solution make you tied to a single manufacturer? | X | Changing the platform at your own request unlikely. | 3 (2,3%) boolean |